\documentclass[aps,preprintnumbers,showpacs,nofootinbib]{revtex4}
\usepackage[english]{babel}
\usepackage{graphicx}
\usepackage{subfig}
\usepackage{hyperref}
\usepackage{dcolumn}
\usepackage{bm}
\newcommand{\newc}{\newcommand}
\newc{\ra}{\rightarrow}
\newc{\lra}{\leftrightarrow}
\newc{\beq}{\begin{equation}}
\newc{\eeq}{\end{equation}}
\newc{\barr}{\begin{eqnarray}}
\newc{\earr}{\end{eqnarray}}

\def\barr{\begin{eqnarray}}
\def\earr{\end{eqnarray}}

\begin{document}
\newcommand{\Od}{{\cal O}}
\newcommand{\lsim}   {\mathrel{\mathop{\kern 0pt \rlap
  {\raise.2ex\hbox{$<$}}}
  \lower.9ex\hbox{\kern-.190em $\sim$}}}
\newcommand{\gsim}   {\mathrel{\mathop{\kern 0pt \rlap
  {\raise.2ex\hbox{$>$}}}
  \lower.9ex\hbox{\kern-.190em $\sim$}}}

\title{
 Theoretical direct WIMP detection rates for transitions to excited states.}

\author{J.D. Vergados$^{1}$, H. Ejiri$^2$ and  K. G. Savvidy$^3$ }
\affiliation{$^1$ Theoretical Physics,University of Ioannina, Ioannina, Gr 451 10, Greece}
\affiliation{$^2$RCNP, Osaka University, Osaka, 567-0047, Japan and\\
  Nuclear Science, Czech Technical University, Brehova, Prague, Czech Republic.}
\affiliation{ $^3$ Department of Physics, Nanjing University, Hankou Lu 22, Nanjing, 210098, China}
\vspace{0.5cm}
\begin{abstract}
The recent WMAP and Planck data have confirmed that exotic dark matter
together with the vacuum energy (cosmological constant) dominate
in the flat Universe. Many extensions of the standard model provide dark matter candidates, in particular Weakly Interacting Massive Particles (WIMPs).
 Thus the direct dark matter detection is central to
particle physics and cosmology.  Most of the research on this
issue has hitherto focused on the detection of the recoiling
nucleus. In this paper we study transitions to the excited states, possible in some nuclei, which have sufficiently low lying excited states. Good examples are the first excited states of $^{127}$I and $^{129}$Xe. 
We find appreciable branching ratios  for the inelastic scattering mediated by the spin cross sections.
So, in principle,
the extra signature of the gamma ray following the de-excitation of these states can, in principle,
be exploited experimentally.
\end{abstract}
\date{\today}
\maketitle
\section{Introduction}
The combined MAXIMA-1 \cite{MAXIMA-1}, BOOMERANG \cite{BOOMERANG},
DASI \cite{DASI} and COBE/DMR Cosmic Microwave Background (CMB)
observations \cite{COBE} imply that the Universe is flat
\cite{flat01}
and that most of the matter in
the Universe is Dark \cite{SPERGEL},  i.e. exotic. These results have been confirmed and improved
by the recent WMAP  \cite{WMAP06} and Planck \cite{PlanckCP13} data. Combining 
the data of these quite precise measurements one finds:
$$\Omega_b=0.0456 \pm 0.0015, \quad \Omega _{\mbox{{\tiny CDM}}}=0.228 \pm 0.013 , \quad \Omega_{\Lambda}= 0.726 \pm 0.015~$$
(the more  recent Planck data yield a slightly different combination $ \Omega _{\mbox{{\tiny CDM}}}=0.274 \pm 0.020 , \quad \Omega_{\Lambda}= 0.686 \pm 0.020)$. It is worth mentioning that both the WMAP and the Plank observations yield essentially the same value of $\Omega_m h^2$,
  but they differ in the value of $h$, namely $h=0.704\pm0.013$ (WMAP) and $h=0.673\pm0.012$ (Planck).
Since any ``invisible" non exotic component cannot possibly exceed $40\%$ of the above $ \Omega _{\mbox{{\tiny CDM}}}$
~\cite {Benne}, exotic (non baryonic) matter is required and there is room for cold dark matter candidates or WIMPs (Weakly Interacting Massive Particles).\\
Even though there exists firm indirect evidence for a halo of dark matter
in galaxies from the
observed rotational curves, see e.g. the review \cite{UK01}, it is essential to directly
detect such matter in order to 
unravel the nature of the constituents of dark matter. At present there exist many such candidates: the
LSP (Lightest Supersymmetric Particle) \cite{ref2a,ref2b,ref2c,ref2,ELLROSZ,Gomez,ELLFOR}, technibaryon \cite{Nussinov92,GKS06}, mirror matter\cite{FLV72,Foot11}, Kaluza-Klein models with universal extra dimensions\cite{ST02a,OikVerMou} etc. Additional theoretical tools are the structure of the nucleus see e.g.  \cite{JDV06a,Dree00,Dree,Chen}, and the nuclear matrix elements \cite{Ress,DIVA00,JDV03,JDV04,VF07}.

In most
calculations the WIMP is supposed to be the  neutralino or LSP (lightest supersymmetric particle), which is assumed to be primarily a gaugino,
usually a bino. Models which predict a substantial fraction of
higgsino lead to a relatively large spin induced cross section due
to the Z-exchange. Such models tend to violate the LSP relic
abundance constraint and are not favored.  Some claims have
recently been made, however, to the effect that the WMAP relic
abundance constraint can be satisfied in the hyperbolic branch of
the allowed SUSY parameter space, even though the neutralino is
then primarily a higgsino \cite{CCN03}. We will not restrict ourselves in supersymmetry and we will  adopt the optimistic view that
the detection rates due to the spin may be large enough to be
exploited by the experiments, see, e.g., \cite{CCN03}
\cite{CHATTO}, \cite{WELLS}, \cite{JDV03} . Such a view is further
encouraged by the fact that, unlike the scalar interaction, the
axial current allows one to populate excited nuclear states,
provided that their energies are sufficiently low so that they are
accessible by the low energy LSP, a prospect proposed long time
ago \cite{GOODWIT} and considered in some detail by Ejiri and collaborators \cite{EFO93}. For a Maxwell-Boltzmann (M-B) velocity distribution 
the average kinetic energy of the WIMP is:
 \beq \langle T\rangle \approx50~\mbox{keV}
\frac{m_{\chi}}{100~\mbox{GeV}}
 \label{kinen}
 \eeq
 So for sufficiently heavy WIMPs the
average energy may exceed the excitation energy, e.g. of 
$57.7~$keV for the $7/2^{+}$ excited state of $^{127}I$ and 39.6 keV for the first excited $3/2^{+}$ of $^{129}$Xe. These are nuclei employed as targets in the ongoing dark matte searches. In other words one
can explore the high velocity window, up to the escape velocity of
$\upsilon_{esc}\approx 550-620~km/s$. From a Nuclear Physics point of
view this transition is not expected  to be suppressed, since they appear to be allowed Gamow-Teller like.

\section{The structure of the nuclei I-127 and Xe-129}
\begin{figure}
\begin{center}
\subfloat[]
{
\includegraphics[width=.5\textwidth]{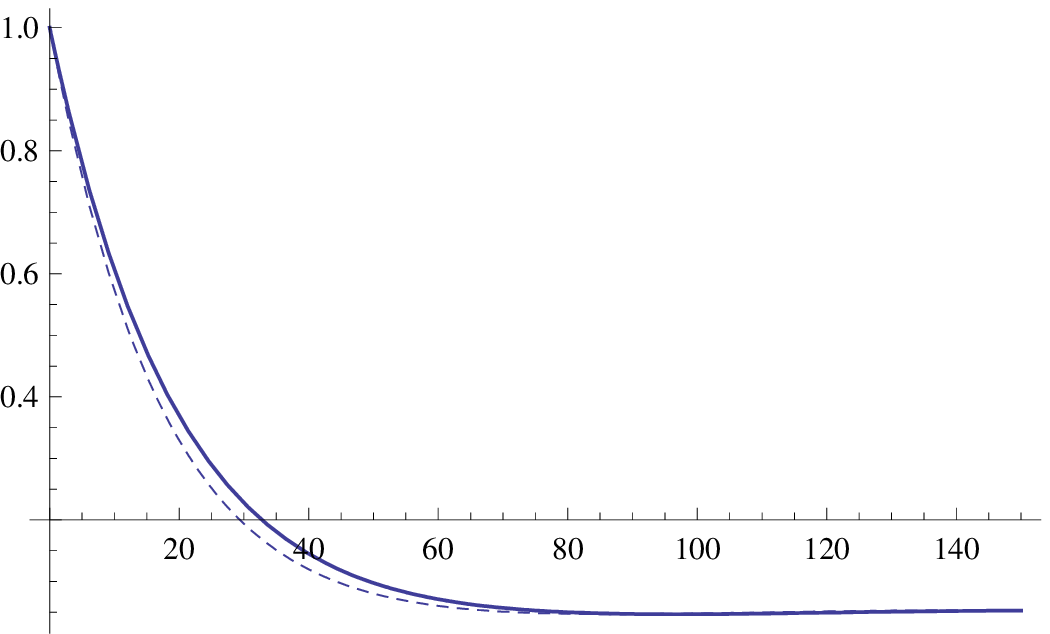}
}\\
{\hspace{-1.5cm} $E_R\rightarrow$keV}
\subfloat[]
{
\includegraphics[width=.5\textwidth]{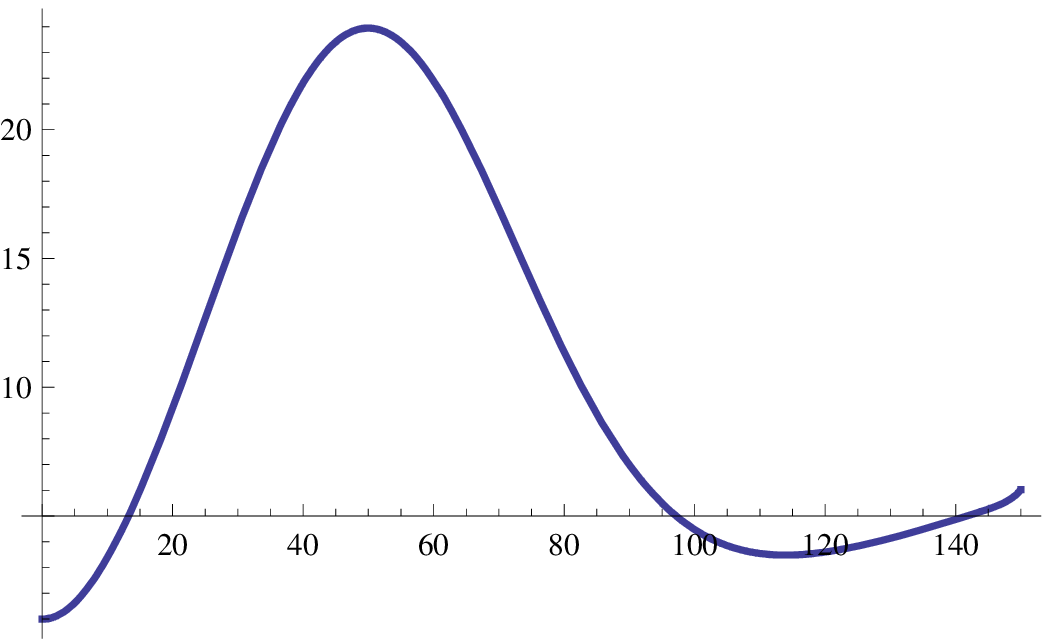}
}
\subfloat[]
{
\includegraphics[width=.5\textwidth]{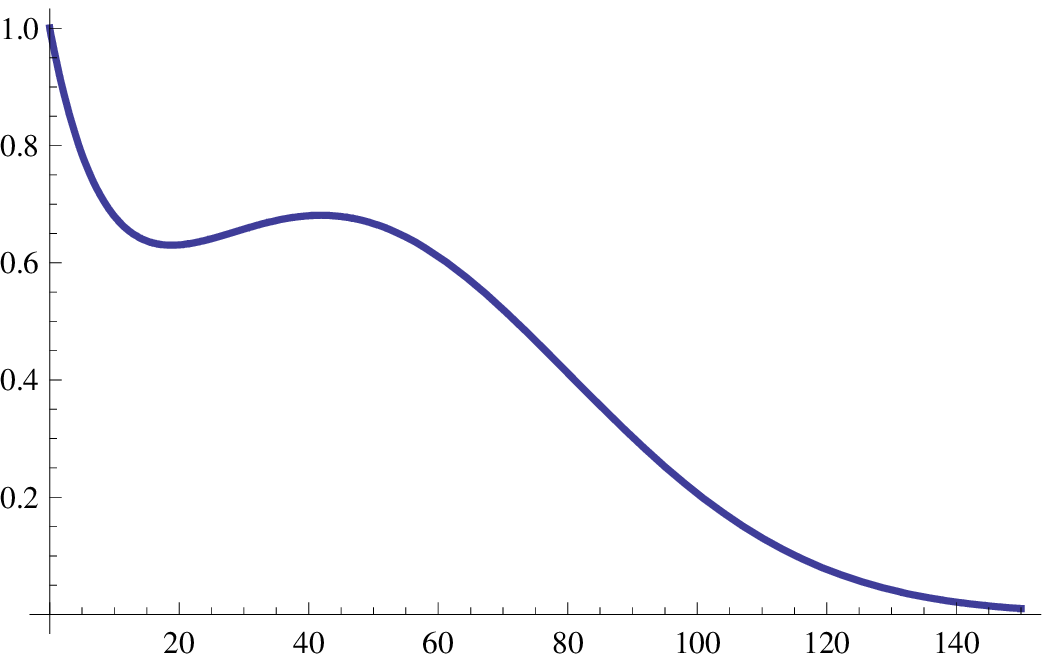}
}
\\
{\hspace{-1.5cm} $E_R\rightarrow$keV}
\caption{ (a) The spin structure function, solid line, and the square of the form factor  for $^{127}$I as a function of the energy transfer in keV involving the ground sate of $^{127}$I,  and the spin structure function related to the excited state of (b) of $^{127}$I and (c) as in (b) for the $^{129}$Xe target.
 \label{fig:FFspin}}
\end{center}
\end{figure}

As it has already been mentioned  these nuclei are
popular targets for dark matter detection.  As a result the
structure of their ground states has been studied theoretically by a
lot of groups. 
\begin{itemize}
\item The $^{127}$  nucleus.\\
This is the most studied case. In this case we mention again the work
 of Ressel and Dean \cite{Ress}, the work of Engel, Pittel and Vogel
\cite{PITTEL94},Vogel and Engel \cite{VogEng},
 Iachello, Krauss and Maiano \cite {IACHELLO91}, 
Nikolaev and Klapdor-Kleingrothaus \cite{NIKOLAEV93}
 and later by Suhonen and collaborators
\cite{SUHONEN03}. 
In all these calculations  for $^{127}$I it appears that the
spin matrix element is dominated by its proton component, which in
our notation implies that the isoscalar and the isovector
components are the same. In these calculations there appears to be
a spread in the spin matrix elements ranging from $0.07$ up to
$0.354$, in the notation of Ressel and Dean \cite{Ress}. This, of
course, implies discrepancies of about a factor of 25 in the event
rates. 
 Furthermore in the context of deformed nuclei \cite{VQS04} for the elastic case one finds:
 \beq
\Omega_0^2= \Omega_1^2 =\Omega_0 \Omega_1=0.164,
 \label{omegas0}
 \eeq
 which is also smaller than the  recent result \cite{SUHONEN03}, while 
for the transition to the excited state this model yields:
 \beq
\Omega_0^2= \Omega_1^2 =\Omega_0 \Omega_1=0.312
 \label{omegas1}
 \eeq
 In yet another calculation \cite{Nsuhonen} in the case of the A=127 system it is reported that :
 $$\Omega _0=1.001,\, \Omega _1=0.868 \mbox{ (elastic)},\quad
 \Omega _0=0.098,\, \Omega _1=0.066 \mbox{ (inelastic)}$$
 \item  Realistic calculations in the case of  the Xe isotopes.\\
  Such calculations relevant for elastic scattering have  recently appeared \cite{MeGazSCH12}. For both elastic and inelastic scattering  the above mentioned  results \cite{Nsuhonen} yield
 for  $^{129}$Xe :
  $$\Omega _0=0.941,\, \Omega _1=-0.954 \mbox{ (elastic)},\quad \Omega _0=0.306,\, \Omega _1=-0.311 \mbox { (inelastic)}$$
 Their spin structure functions are presented in Fig. \ref{fig:FFspin}.\\
 In the present calculation we are going to employ the last two sets of spin nuclear  matrix elements 
 (a summary of some nuclear ME involved in elastic scattering can be found elsewhere \cite{SavVer13}).
 \end{itemize}

\section{The formalism for the WIMP-nucleus differential event rate}
The expression for the elastic differential event rate is well known, see e.g. \cite{SavVer13}. The time averaged rate  can be cast in the form:
\beq
\left .\frac{d R_0}{ dE_R }\right |_A=\frac{\rho_{\chi}}{m_{\chi}}\,\frac{m_t}{A m_p}\, \left (\frac{\mu_r}{\mu_p} \right )^2\, \sqrt{<\upsilon^2>} \,\frac{1}{Q_0(A)} A^2 \sigma_N^{\mbox{\tiny{coh}}}\left .\left (\frac{d t}{du}\right ) \right |_{\mbox{ \tiny coh}} 
\label{drdu}
\eeq
with with $\mu_r$ ($\mu_p$) the WIMP-nucleus (nucleon) reduced mass and $A$ is the nuclear mass number. $ m_{\chi}$ is the WIMP mass, $\rho(\chi)$ is the WIMP density in our vicinity, assumed to be 0.3 GeV cm$^{-3}$,  and $m_t$ the mass of the target. 

Furthermore one can show that
\beq
\left .\left (\frac{d t}{d u}\right )\,\right |_{\mbox{\tiny coh}}=\sqrt{\frac{2}{3}}\, a^2 \,F^2(u)  \, \Psi_0(a \sqrt{u}),\quad \left .\left (\frac{d h}{d u}\right )\right |_{\mbox{\tiny coh}}=\sqrt{\frac{2}{3}}\, a^2 \,F^2(u) \,\Psi_1(a \sqrt{u})
\eeq
The factor $\sqrt{2/3}$ is nothing but $\upsilon_0/\sqrt{\langle \upsilon ^2\rangle}$ since in Eq. (\ref{drdu}) $\sqrt{\langle \upsilon ^2\rangle}$ appears. In the above expressions  $a=(\sqrt{2} \mu_r b \upsilon_0)^{-1}$, $\upsilon_0$ the velocity of the sun around the center of the galaxy and $b$ the nuclear harmonic oscillator size parameter characterizing the nuclear wave function.  $ u$ is the energy transfer $E_R$ in dimensionless units given by
\begin{equation}
 u=\frac{E_R}{Q_0(A)}~~,~~Q_{0}(A)=[m_pAb^2]^{-1}=40A^{-4/3}\mbox{ MeV}
\label{defineu}
\end{equation}
and $F(u)$ is the nuclear form factor. Note that the parameter $a$ depends both on the WIMP mass, the target and the velocity distribution. Note also that for a given energy transfer $E_R$ the quantity $u$ depends on $A$. 

For the axial current (spin induced) contribution
 one finds:
\beq
\left .\frac{d R_0}{ dE_R }\right |_A=\frac{\rho_{\chi}}{m_{\chi}}\,\frac{m_t}{A m_p}\, \left (\frac{\mu_r}{\mu_p} \right )^2\, \sqrt{<\upsilon^2>} \,\frac{1}{Q_0(A)} \frac{1}{3}\,\left ( \Omega_p-\Omega_n\right )^2\, \sigma_N^{\mbox{\tiny{spin}}}\left .\left (\frac{d t}{du}\right ) \right |_{\mbox{ \tiny spin}}
\label{drdus}
\eeq
with
\beq
\left . \left (\frac{d t}{d u}\right )\right |_{\mbox{\tiny spin}}=\sqrt{\frac{2}{3}} \, a^2  \, F_{11}(u)   \,  \Psi_0(a \sqrt{u}),\quad \left .\left (\frac{d h}{d u}\right )\right |_{\mbox{\tiny spin}}=\sqrt{\frac{2}{3}}  \, a^2  \, F_{11}(u)  \, \Psi_1(a \sqrt{u}),
\eeq
where $F_{11}$ is the spin response function (the square of the spin form factor). The behavior of the spin responce function $F_{11}$ for the isovector (isospin 1) channel is exhibited in Fig. \ref{fig:FFspin}. The other spin response functions $F_{01}$ and $F_{00}$ are, in our normalization, almost identical to the one shown. Note that, in the cases shown in Fig. \ref{fig:FFspin}, the spin response functions are not different from the square of the form factors entering the coherent mode. Thus, except for the scale of the event rates, the behavior of the coherent and the spin modes is almost identical.

Integrating the above differential rates we obtain the total rate including the time averaged rate  and the relative modulation amplitude $h$ for each mode  given by:
\beq
R_{\mbox{\tiny coh}}=\frac{\rho_{\chi}}{m_{\chi}} \, \frac{m_t}{A m_p}  \left ( \frac{\mu_r}{\mu_p} \right )^2  \sqrt{<\upsilon^2>} \, A^2 \, \sigma_N^{\mbox{\tiny{coh}}} \, t_{\mbox{\tiny coh}},\quad t_{\mbox{\tiny coh}}=\int_{E_{th}/Q_0(A)}^{(y_{\mbox{\tiny esc}}/a)^2} \, \left .\frac{dt}{du}\right |_{\mbox{\tiny coh}}du
\label{Eq:Trates}
\eeq
\beq
R_{\mbox{\tiny spin}}=\frac{\rho_{\chi}}{m_{\chi}} \, \frac{m_t}{A m_p} \, \left ( \frac{\mu_r}{\mu_p} \right )^2  \sqrt{<\upsilon^2>} \, \frac{1}{3} \, \left ( \Omega_p-\Omega_n\right )^2 \, \sigma_N^{\mbox{\tiny{spin}}} \, t_{\mbox{\tiny spin}} \, \quad t_{\mbox{\tiny spin}}=\int_{E_{th}/Q_0(A)}^{(y_{\mbox{\tiny esc}}/a)^2} \, \left .\frac{dt}{du}\right |_{\mbox{\tiny spin}}du
\label{Eq:Tratec}
\eeq
for each mode (spin and coherent). $E_{th}(A)$ is the energy threshold imposed by the detector.

Using the expressions for nucleon cross sections, (\ref{Eq:Trates}) and (\ref{Eq:Tratec}), we can obtain the total rates. These expressions contain the following parts: i) the parameter $t$, which contain the effect of the velocity distribution and the nuclear form factors ii) the elementary nucleon cross sections iii) the nuclear physics input (nuclear spin spin ME as mentioned above.\\
\section{Some results}
For purposes of illustration we will employ the  nucleon cross sections obtained in a recent work \cite{SavVer13}, without committing ourselves to that particular model. From expessions (\ref{Eq:Trates}) and (\ref{Eq:Tratec}), we can obtain the total rates. These expressions contain the following parts: i) the parameter $t$ , which contain the effect of the velocity distribution and the nuclear form factors ii) the elementary nucleon cross sections iii) the nuclear physics input (nuclear spin spin ME)
\\ We have seen that in the model we are going to employ \cite{SavVer13} for demonstrating purposes,   there is no $A^2$ or $Z^2$ coherence and the isovector spin induced cross section is the only possibility.
 \subsection{The differential event rates} 
\begin{figure}
\begin{center}
\subfloat[]
{
\rotatebox{90}{\hspace{0.0cm} $\left .\left (\frac{dR_0}{dE_R}\right )\right|_A \rightarrow$(kg-y)/ keV}
\includegraphics[height=.25\textwidth]{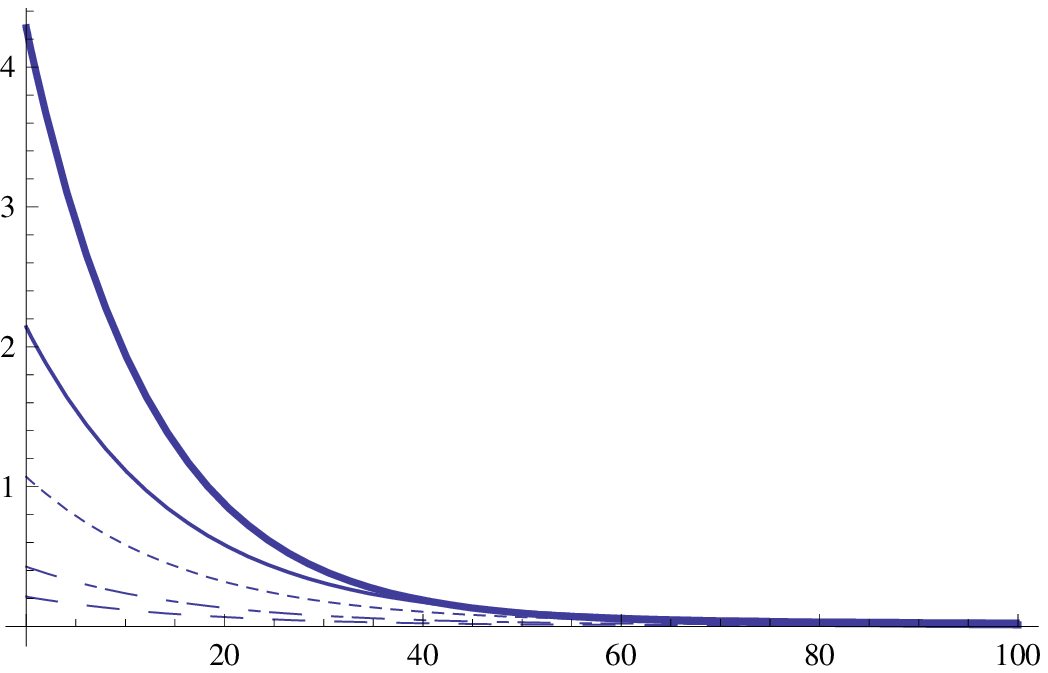}
}
\subfloat[]
{
\rotatebox{90}{\hspace{0.0cm} $\left .\left (\frac{dR_0}{dE_R}\right )\right|_A\rightarrow$(kg-y)/ keV)}
\includegraphics[height=.25\textwidth]{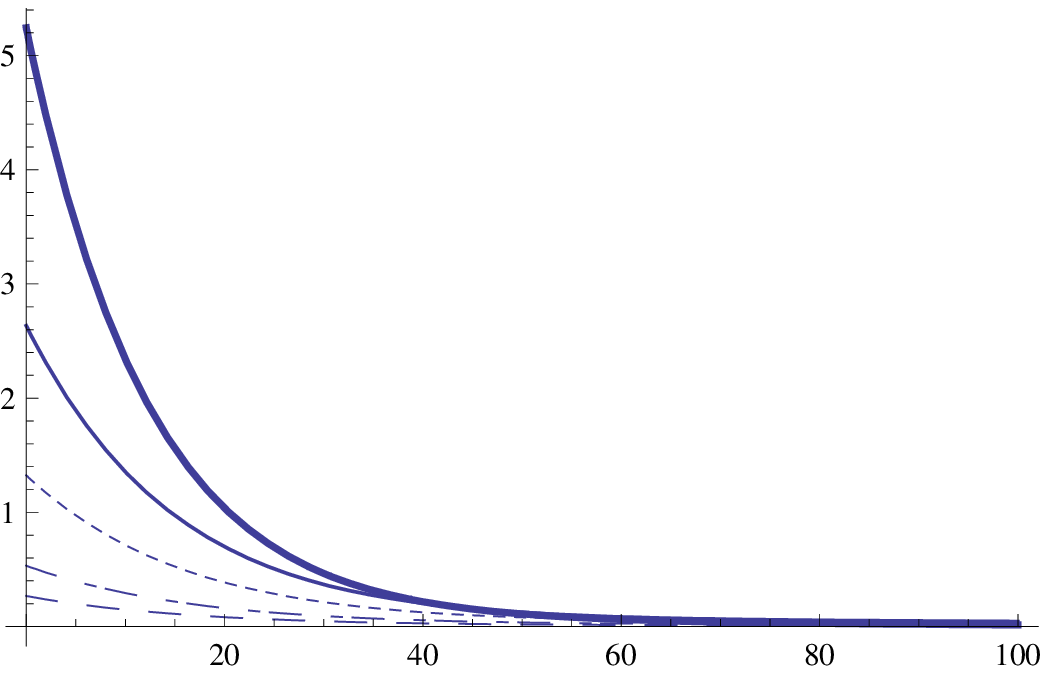}
}
\\
{\hspace{-2.0cm} $E_R\rightarrow$keV}\\
\caption{ We show the time average differential rate  $\left .\frac{dR_0}{dE_R}\right|_A$,  as a function of the recoil energy $E_R$  in keV for elastic scattering. These results correspond to the spin mode in the case of $^{127}$I (a) and $^{129}$Xe (b). The graphs from top to bottom correspond to WIMP masses (50, 100,  200, 500 , 1000) GeV. The escape velocity was taken to be $\upsilon_{esc}=2.8 \upsilon_0.$ The effect of quenching has not been included.
 \label{fig:dRhdQS_127}}
\end{center}
\end{figure}

The differential event rates, perhaps the most  interesting from an experimental point of view,  depend on the WIMP mass.  So we can only present them for some select masses. Considerations based on the relic abundance of the WIMP in this model lead to the conclusion that it has a  mass between 80 and 200 GeV. This the WIMP mass range of interest to us.  For illustration purposes, however, we have decided to present some  results also for  lighter WIMPs. Our results for the  differential rates are  exhibited    in Fig. \ref{fig:dRhdQS_127}.
 \subsection{The total event rates}

%

\begin{figure}
\begin{center}
\subfloat[]
{
\rotatebox{90}{\hspace{0.0cm}$R_{\mbox{\tiny spin}}\rightarrow$ events/kg-y}
\includegraphics[width=0.4\textwidth]{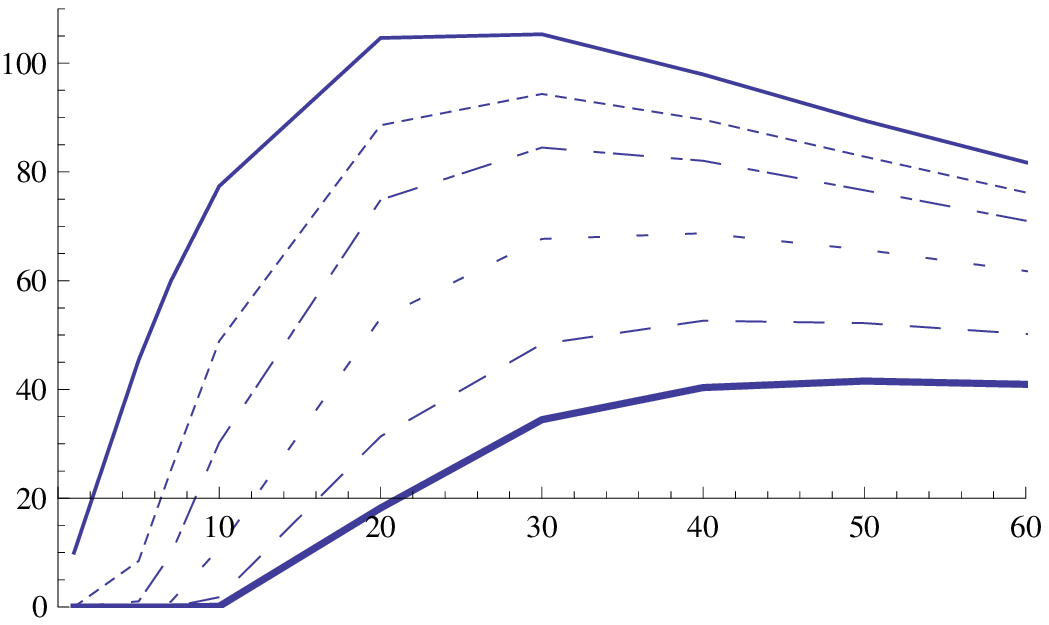}
}
\subfloat[]
{
\rotatebox{90}{\hspace{0.0cm} $R_{\mbox{\tiny spin}}\rightarrow$events/kg-y}
\includegraphics[width=0.4\textwidth]{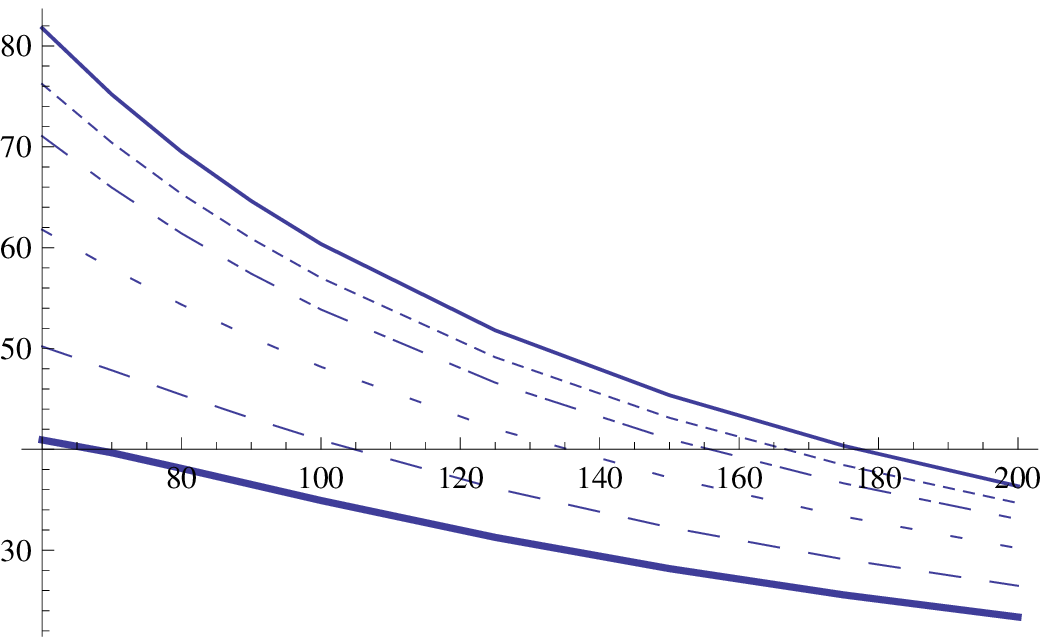}
}
\\
{\hspace{-2.0cm} $m_{\chi} \rightarrow$GeV}
\caption{ The  predicted time averaged total rate $R_{\mbox{\tiny spin}}$ for $^{127}$I  restricted to the small WIMP mass regime  (a) and its restriction to the   WIMP mass range  relevant to our model (b). The highest curve corresponds to zero threshold and the lowest to a threshold of 10 keV. The effect of quenching fas not been included.
 \label{fig:plotRs127}} 
\end{center}
\end{figure}

 We are interested here in the spin induced rates become relevant. Such results for the time averaged total rate are shown in Figs \ref{fig:plotRs127}. The obtained results in the case of $^{129}$Xe are a bit different due to the somewhat different static spin ME.
\section{transitions to excited states}
Transitions to excited states are normally forbidden, except in some odd nuclei that have low lying excited states. Then transitions to excited states are possible due to the spin. They most favorable are expected to be those that based on the total angular momentum of the states involved appear to be  Gamow-Teller like transitions. A good possibility is  the $7/2^+$ state of $^{127}$I,  which is at $57.6$ keV above the $5/2^+$ ground state, and the the $3/2^+$ state of $^{129}$Xe at 39.6 keV. Inelastic transitions look like a Gammow-Teller like, $\Delta J=1$, but, unfortunately, the dominant component of the wave function is not $\Delta \ell=0$. So the spin ME entering the inelastic component may be suppressed.
\subsection{Elementary considerations}
The first criterion is to have a large elementary nucleon cross section. On general grounds, which have to do with the transition from the quark level, where particle calculations are involved to the nucleon level, only the isovector amplitude is important. There exist particle models that predict such large cross sections. Since, however, the accessible excited states involve targets with large A, the branching ratio to the excited state is going to be significant only if the coherent nucleon cross section happens to be small. There exist particle models that satisfy this criterion. The WIMP spin 3/2 model discussed above \cite{SavVer13} satisfies this criterion. In fact this WIMP is expected to be a Majorana Fermion, in which case the (neutron) coherent event rate vanishes. Another possibility is in supersymmetry in the co-annihilation region \cite{Cannoni11}, where the ratio of the spin to coherent nucleon cross section, depending on $\tan{\beta}$ and the WIMP mass, which is in the range 200-500 GeV,  can be as large as $10^3$. In a region of the model space the ratio of the elastic spin cross section to the coherent can be as large as 10$\%$. More recent calculations in the supersymmetric $SO(10)$ model \cite{Gogoladze13}, also in the co-annihilation region, predict large spin to coherent cross section ratio, of the order of $2\times 10^{3}$  and a WIMP mass of about 850 GeV.
Thus from the particle physics point of view the prospect of getting an appreciable branching ratio is is not discouraging.
\subsection{Isotope considerations}
Transitions to excited states are normally forbidden, except in some odd nuclei that have low lying excited states with $E \le 100 $ keV, because WIMP energy is mostly below 150 KeV. Then transitions to excited states are possible due to the spin interaction with SD WIMPs. 

Possible odd nuclei involved in DM detectors used for WIMP searches are the 57.6 keV state in $^{127}$I and the 39.6 keV state in $^{129}$Xe, as given in Table \ref{tab:ej1}. The spin excitations of these states are not favored, because the dominant components of the relevant wave functions  are characterized by $\Delta \ell\ne 0$, i.e. $\ell$ forbidden transitions.  Nevertheless the spin transitions are possible, due to the small components  as seen from the M1 $\gamma $ transition rates.

Experimentally, large scale I and Xe detectors are being used at several DM collaborations with natural isotopes. The odd isotope abundance ratios are 100 $\%$ and 26$\%$ for $^{127}$I and $^{129}$Xe, respectively. Thus it is quite realistic to study the inelastic excitations in these nuclei to search for SD WIMPs. 

In fact experimental observation of the inelastic excitation has several advantageous points. The experimental detection is discussed in section VI.

\begin{table} [h]
\caption{Inelastic spin excitations of $^{127}$I and $^{129}$Xe. $A$ : natural abundance ratio, $E$ : excitation energy, $J_i$: ground state spin parity, $J_f$ : excited state spin parity, and  $T_{1/2 } $: half life.}
\label{tab:ej1}
\begin{center}
\begin{tabular}{cccccc} 
  \\ 
\hline
Isotope &   $A(\%)$ &   $E$ (keV) &   $J_i$ &   $J_f$ &  $T_{1/2} $(n sec) \\        
\hline
$^{127}$I    &       100  &        57.6  &    5/2$^+$  &  7/2$^+$ &    1.9 \\  
$^{129}$Xe &       26.4  &      39.6  &    1/2$^+$  &   3/2$^+$ &     0.97\\     

 \hline
\end{tabular}
\end{center}
\medskip 
\end{table}  

\subsection{Kinematics}

\begin{figure}
\begin{center}
\subfloat[]
{
\rotatebox{90}{\hspace{0.0cm} $\left .\left (\frac{dR_0}{dE_R}\right )\right|_A \rightarrow$(kg-y)/ keV}
\includegraphics[height=.25\textwidth]{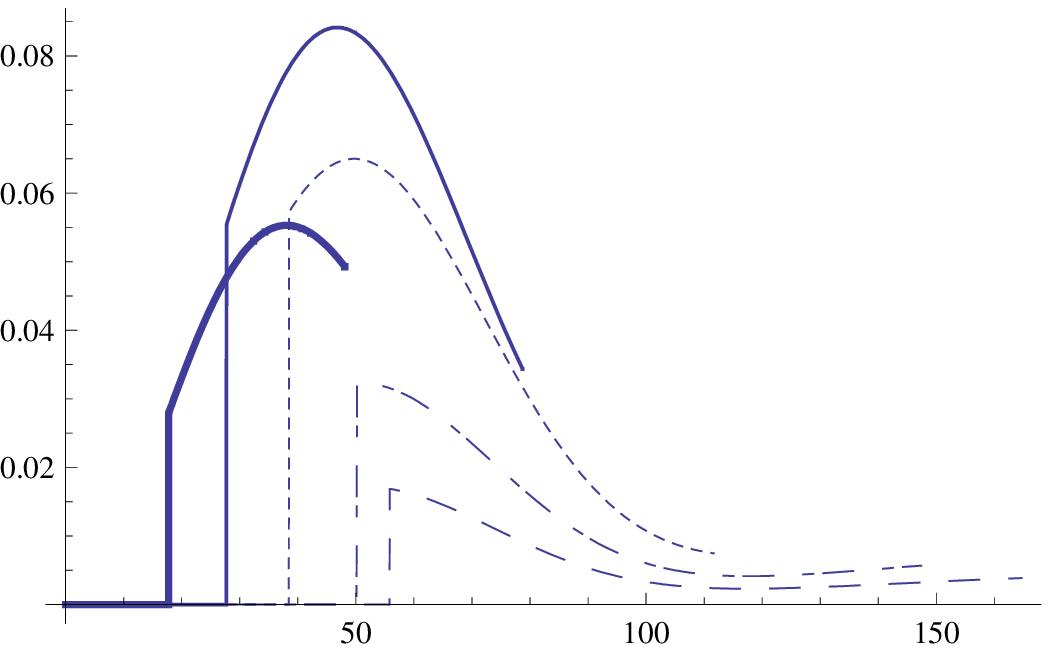}
}
\subfloat[]
{
\rotatebox{90}{\hspace{0.0cm} $\left .\left (\frac{dR_0}{dE_R}\right )\right|_A\rightarrow$(kg-y)/ keV)}
\includegraphics[height=.25\textwidth]{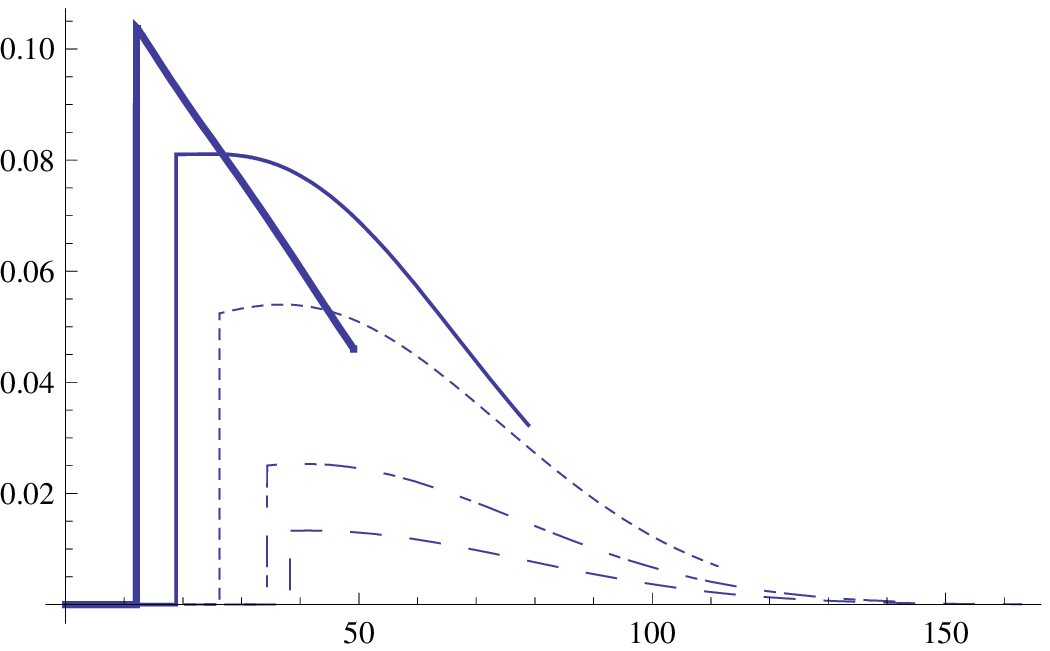}
}
\\
{\hspace{-2.0cm} $E_R\rightarrow$keV}\\
\caption{ The same as in Fig. \ref{fig:dRhdQS_127} in the case of the inelastic scattering.
 \label{fig:dRhdQExc_127}}
\end{center}
\end{figure}

The evaluation of the differential rate for the inelastic transition proceeds as in the elastic discussed above except:
\begin{enumerate}
\item The transition spin matrix element must be used.
\item The transition spin response function must be used.\\
For Gammow teller like transitions, it does not vanish a zero energy transfer. So it can be normalized to one, if the static spin value is taken out of the ME.
\item The kinematics is modified.\\
 The energy-momentum conservation reads:
 \beq
 \frac{-q^2}{2 \mu_r}+\upsilon \xi q -E_x =0, \quad E_x= \mbox {excitation energy}\Leftrightarrow -\frac{m_A}{\mu_r}E_R +\upsilon \xi \sqrt{2 m_A E_R}+E_x =0
 \eeq
 Clearly $\xi>0$ as before. Then $\xi<1$  and the reality of $E_R$ impose the conditions:
 \beq
 E_R>E_0,\quad E_0=\frac{\mu_r}{m_A}E_x,\quad \upsilon>\frac{E_x+\frac{m_A}{\mu_r}E_R}{\sqrt{2 m_A E_R}}
 \eeq
 We find it simpler to  deal with the phase space in dimensionless units. Noticing that $u=(1/2) q^2 b^2$ and
 \beq
 \delta \left (\frac{-q^2}{2 \mu_r}+\upsilon \xi q -E_x \right )=\delta \left (-\frac{u}{\mu_r b^2}+\upsilon \xi\frac{\sqrt{2 u}}{b}-E_x \right )\Leftrightarrow \frac{b}{\upsilon \sqrt{2u}}\delta \left (\xi-\frac{E_x+u/(\mu_r b^2)}{\upsilon \sqrt{2u}}\right )
 \eeq
 we find:
 \beq
 \int q^2 d \xi dq \delta \left (\frac{-q^2}{2 \mu_r}+\upsilon \xi q -E_x \right )=\frac{1}{b^2 \upsilon} du
 \eeq
 i.e. we recover the same expression as in the case of ground state transitions.\\
 The above constraints now read:
 \beq
 u>u_0,\quad y>a \frac{u+u_0}{\sqrt{u}}
 \eeq
 \beq
 u_0=\mu_r E_x b^2,\quad a=\frac{1}{\sqrt{2}\mu_r \upsilon_0 b},\quad u=\frac{E_R}{Q_0(A)}
 \eeq
It should be stressed that for transitions to excited states the energy of recoiling nucleus must be above a minimum energy, which depends on the excitation energy and the mass of the nucleus as well as the WIMP mass. This limits the inelastic scattering only for recoiling  energies  above  the values $\left(E_R\right)_{\mbox{min}}$. This explains why the rates do not increase as fast with the WIMP mass as  naively expected. We should also mention that the maximum energy $E_R$ allowed, which is limited by the maximum WIMP velocity, is more constrained in this case compared to that associated to the elastic scattering. In fact
 the maximum energy that can be transferred is:
 \beq
 u_{max}=\frac{1}{4}\left (\frac{y_{esc}}{a}+\sqrt{\left (\frac{y_{esc}}{a} \right )^2-4 u_0^2} \right )^2
 \eeq
 \end{enumerate}
 The maximum energy transfers $ u_{max}$ somewhat depend on the escape velocity. For $y_{esc}=2.8$, which corresponds to $\upsilon_{exc}=620$km/s.
The values $u_0$, $u_{\mbox{max}}$, $\left(E_R\right)_{\mbox{min}}$ and $\left(E_R\right)_{\mbox{max}}$ relevant for the inelastic scattering of $^{127}$I and $^{129}$Xe are shown in Table \ref{tab:tab2}. We see that the energies $\left(E_R\right)_{\mbox{min}}$ are much above threshold.\\
\begin{table}[h]
\caption{ The kinematical parameters entering the inelastic scattering to the first excited state of
$^{127}$I and $^{129}$Xe.}
\label{tab:tab2}
\begin{center}
\begin{tabular}{|l|c|ccccc|}\\
\hline
target& parameter&& $m_{\chi}$(GeV)&\\
\hline
& &50& 100& 200& 500& 1000\\
\hline
$^{127}$I &$u_0$& 0.270& 0.421& 0.585 &0.762& 0.848\\
&$u_{\mbox{max}}$& 0.724& 1.186 &1.683& 2.221& 2.482\\
&$\left(E_R\right)_{\mbox{min}}$(keV)& 17.8& 27.7 &38.5& 50.2& 55.8\\
&$\left(E_R\right)_{\mbox{max}}$(keV)& 47.6& 78.0 &110.7 &146.2& 163.3\\
\hline
$^{129}$Xe &$u_0$& 0.187& 0.293& 0.408& 0.533& 0.594\\
&$u_{\mbox{max}}$& 0.762& 1.227& 1.732 &2.283& 2.551\\
&$\left(E_R\right)_{\mbox{min}}$(keV)& 12.1& 18.9 &26.3& 34.0& 38.3\\
&$\left(E_R\right)_{\mbox{max}}$(keV)& 49.0 &79.1& 111.6 &147.2& 164.4\\
\hline
\end{tabular}
\end{center}
\end{table}
 To avoid uncertainties arising from the relevant particle model, we will present the  rate to the excited relative to that to the ground state (branching ratio).
 The differential event rate for inelastic scattering takes a form similar to the one given by Eq. \ref{drdus} except that 
 $$\Omega_p-\Omega_n\rightarrow \left (\Omega_p-\Omega_n\right )^{\mbox{\tiny{inelastic}}},\, F_{11}(u)\rightarrow F_{11}(u)^{\mbox{\tiny{inelastic}}},\quad\Psi_0(a\sqrt{u})\rightarrow \Psi_0\left(a\frac{ u+u_0}{\sqrt{u}}\right ),\quad u_0\le u\le u_{max}$$ 
 The obtained results are shown in Fig. \ref{fig:dRhdQExc_127}
 \subsection{Branching ratios}
\begin{figure}
\begin{center}
\subfloat[]
{
\includegraphics[width=.4\textwidth]{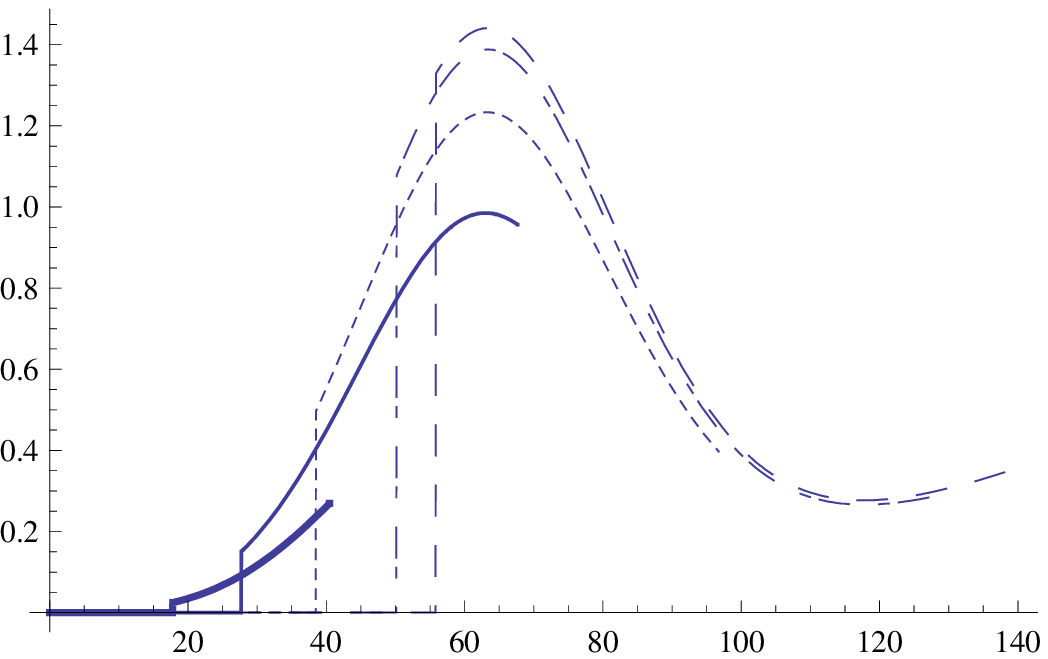}
}
\subfloat[]
{
\includegraphics[width=.4\textwidth]{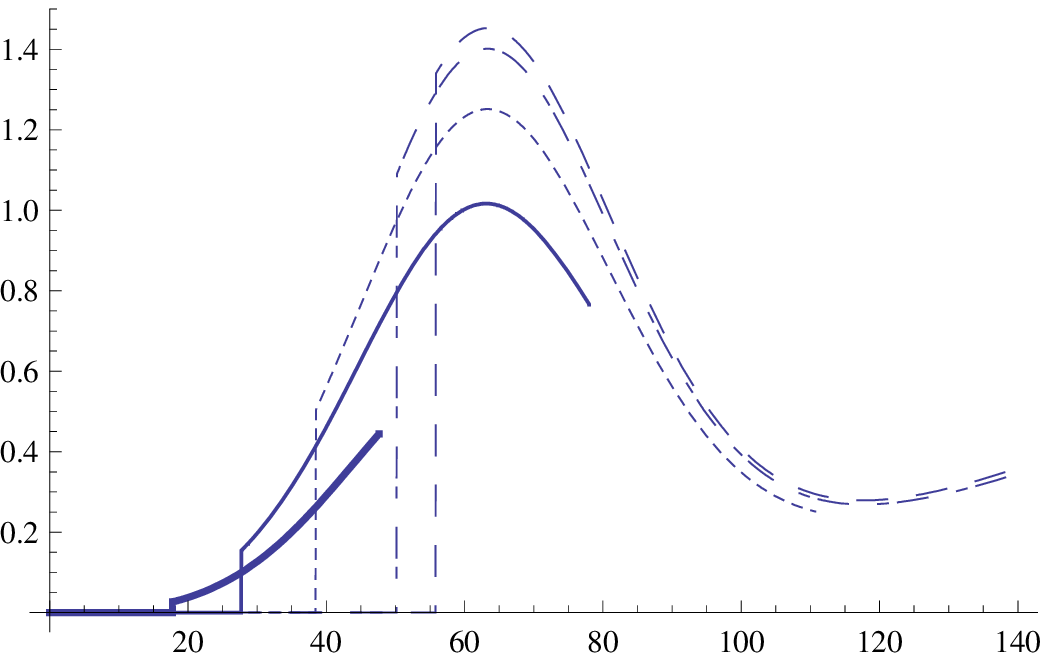}
}\\
{\hspace{-2.0cm} $E_R\rightarrow$keV}
\subfloat[]
{
\includegraphics[width=.4\textwidth]{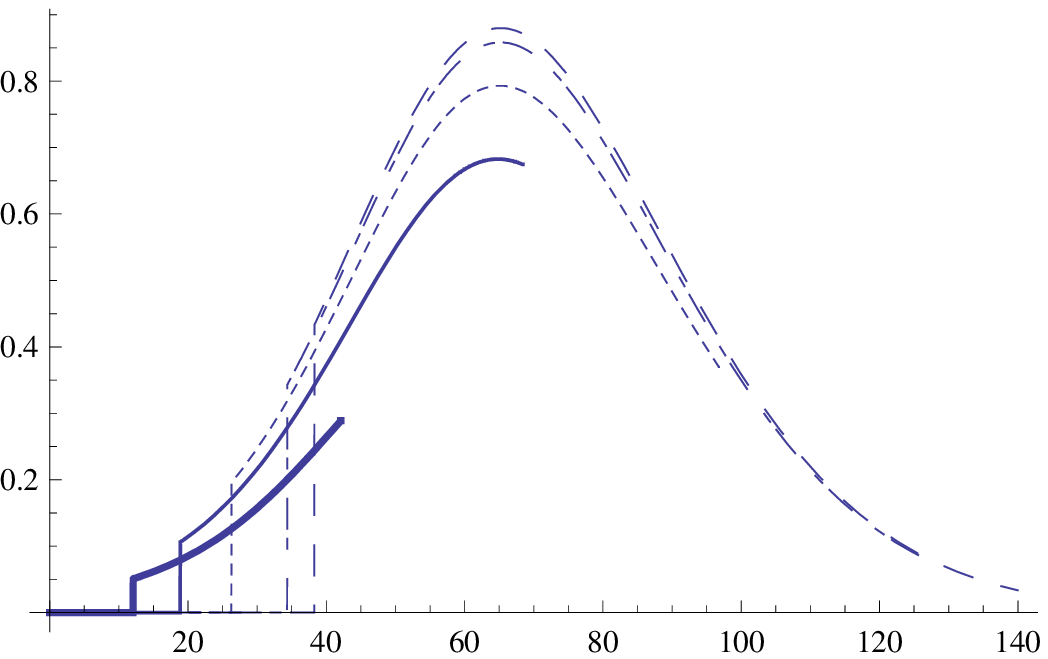}
}
\subfloat[]
{
\includegraphics[width=.4\textwidth]{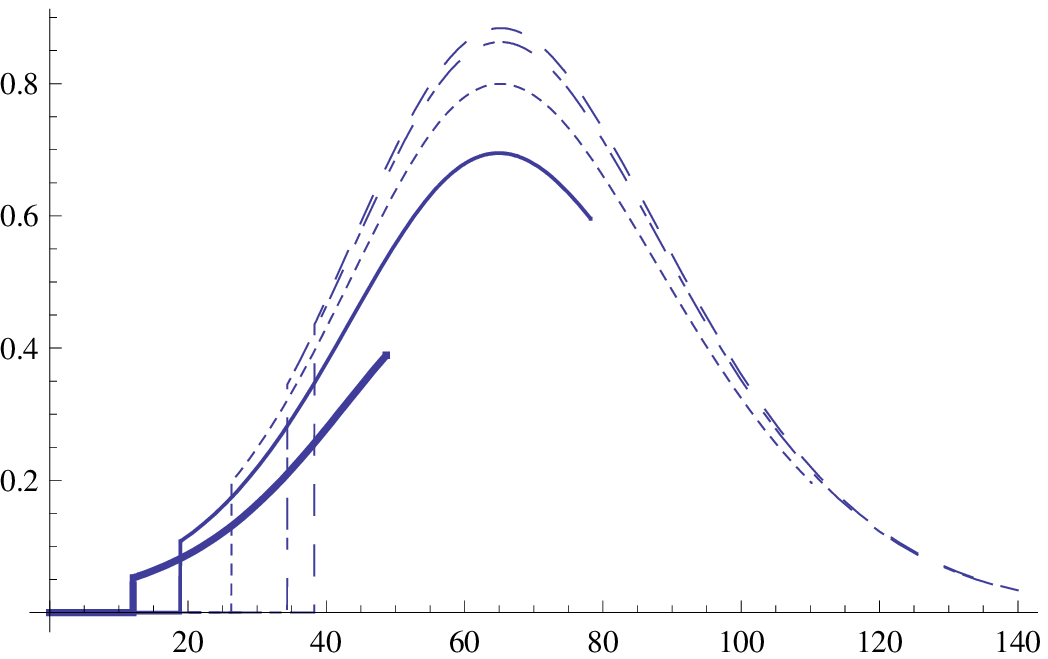}
}\\
{\hspace{-2.0cm} $E_R\rightarrow$keV}
\caption{ The ratio of the differential rates, $\frac{dR(\mbox{excited})}{dE_R}/\frac{dR(\mbox{gs})}{dE_R}$, as a function of the recoil energy $E_R$ in keV. The thick solid, solid, dotted, dashed-dotted and dashed curves correspond to WIMP masses 50, 100, 200, 500 and 1000 GeV. In panel (a) we show the results  for $y_{esc}=2.5$ and in (b) for  rate $y_{esc}=2.8$. In (c) and (d) the same as in (a) and (b) for $^{129}$Xe. Only the spin mode has been taken into account. The effect of quenching has not been included.
 The dependence on the escape velocity in the range of the accepted values is mild.
 \label{fig:difRex}}
\end{center}
\end{figure}

\begin{figure}
\begin{center}
\subfloat[]
{
\includegraphics[width=0.4\textwidth]{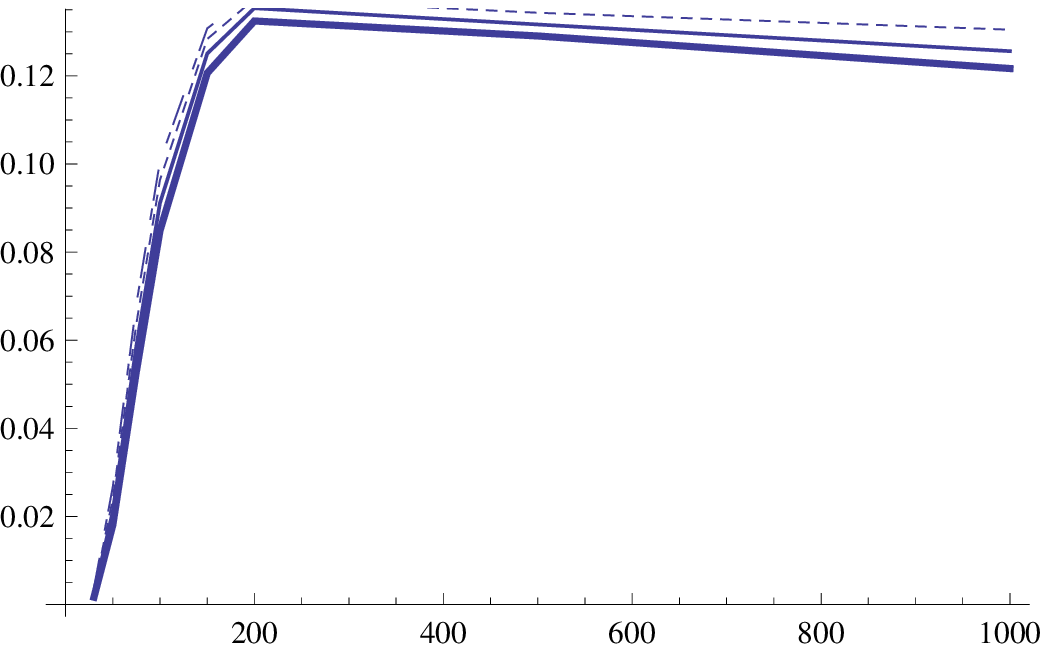}
}
\subfloat[]
{
\includegraphics[width=0.4\textwidth]{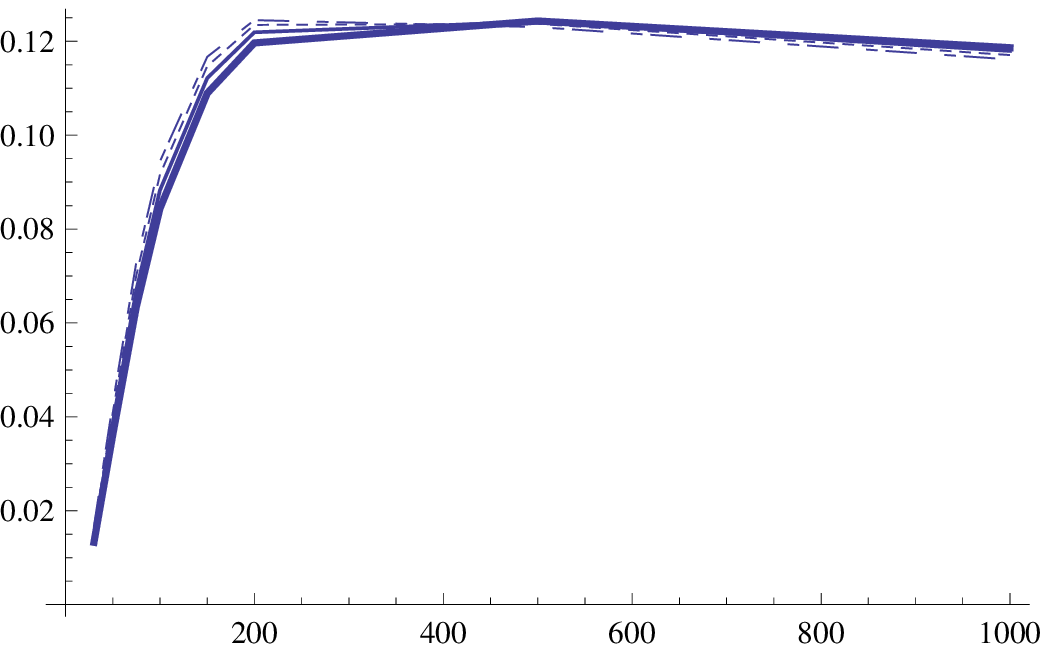}
}\\
{\hspace{-2.0cm} $m_{\chi}\rightarrow$GeV}
\caption{ A sketch of the  ratio of the total rates, $R(\mbox{excited})/R(\mbox{gs})$, as a function of the WIMP mass in GEV for $^{127}$I (a) and $^{129}$Xe (b). The dependence on the escape velocity is not visible. Only the spin mode has been taken into account. The effect of quenching has not been included.
 \label{fig:totRex}}
\end{center}
\end{figure}

\begin{figure}
\begin{center}
\subfloat[]
{
\includegraphics[width=0.4\textwidth]{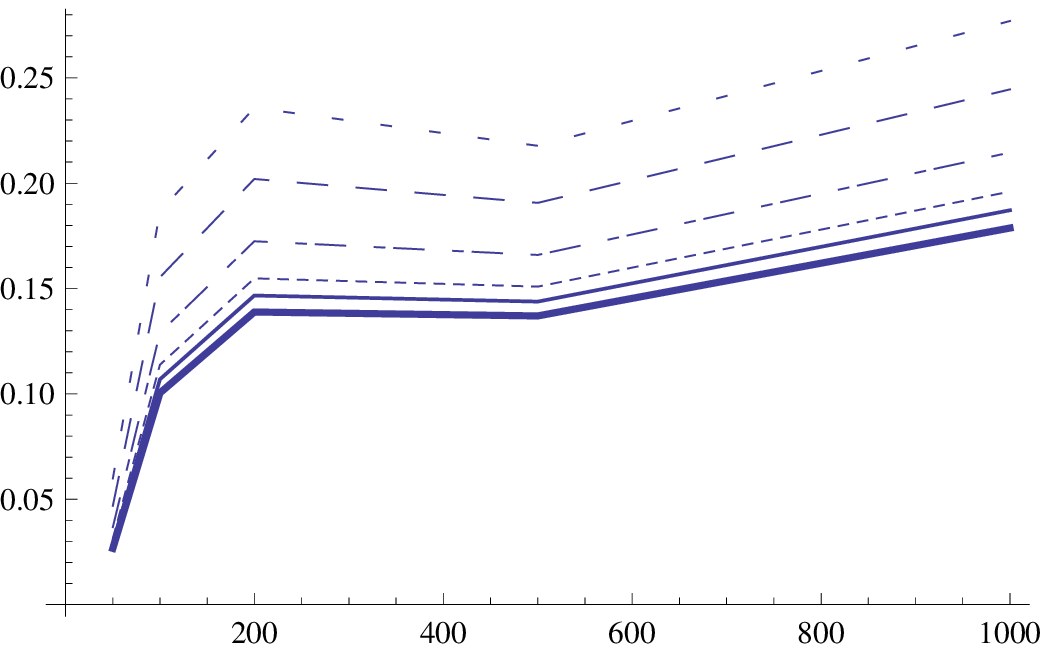}
}
\subfloat[]
{
\includegraphics[width=0.4\textwidth]{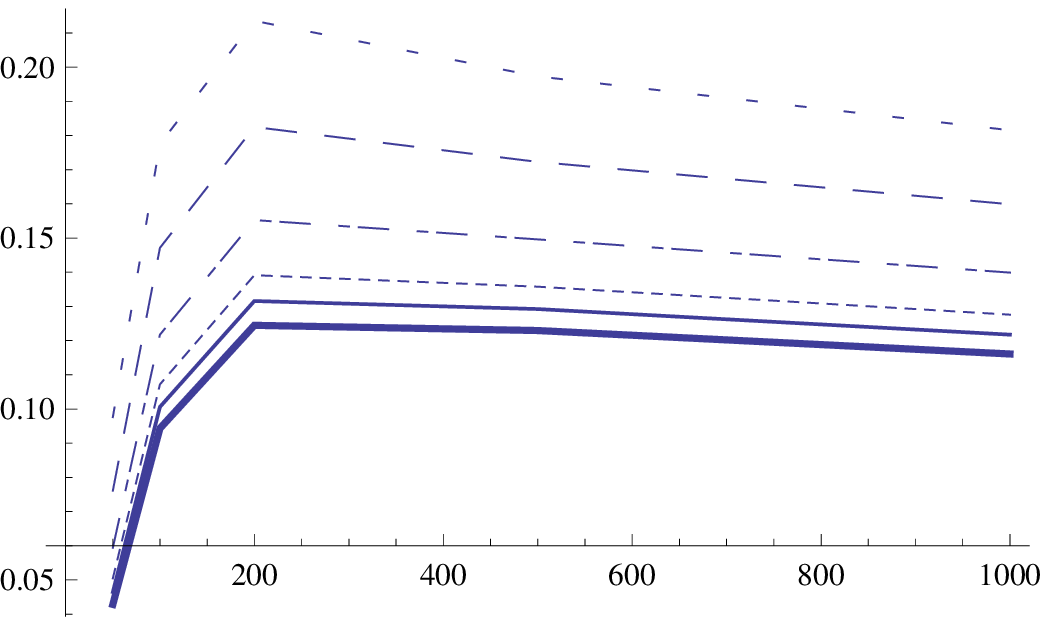}
}\\
{\hspace{-2.0cm} $m_{\chi}\rightarrow$GeV}
\caption{ A sketch of the  ratio of the total rates, $R(\mbox{excited})/R(\mbox{gs})$, as a function of the WIMP mass in GEV for y=2.8 in the case  $^{127}$I (a) and $^{129}$Xe (b) by varying the energy threshold. From bottom up the threshold values are $0,1,2,3,4,7$ and 10 keV. Only the spin mode has been taken into account. The effect of quenching has not been included.
 \label{fig:RexcTh}}
\end{center}
\end{figure}

 We will  evaluate the branching if the inelastic cross section to the spin induced cross section. One may renormalize it appropriately by including the coherent mode since then $\sigma_A^{spin}\rightarrow\sigma_A^{spin}+A^2\sigma_N^{coh}$, with $ \sigma_A^{spin}=(1/3)(\Omega_p-\Omega_n)^2  \sigma_N^{spin}$. As we have mention for the elastic case the form factors for the spin and the coherent mode are almost the same, which means that the shape of the differential rates are similar.
  We will restrict ourselves in the isovector transition. This is the case in the spin 3/2 particle model  and, in general, it is  expected to be dominant \cite{JELLIS93a} due to considerations related to the spin of the nucleon. In the case of $^{127}$I the spin ME to the excited state divided by that involving the ground state is 0.076 \cite{Nsuhonen}. The spin response function (see Fig. \ref{fig:FFspin}), in the region of interest to us favors the excited state, which compensates for the smallness of the static spin value. In the case of $^{129}$Xe this ratio 0.326 but the spin structure function is not so favorable, see Fig. \ref{fig:FFspin}.
Our results for the differential rate are shown in Fig. \ref{fig:difRex}.
The individual differential rates, are, of course, suppressed at high energy transfer due to the nuclear form factor, which pretty much cancels in the ratio.\\ Under the same assumptions the branching ratio, i.e. the ratio of the total rates, is sketched in fig. \ref{fig:totRex}.
\\Since the elastic event rate is reduced by the threshold effects, but the inelastic transition is not affected by such effects, we expect the branching ratio to be increasing as the threshold energy is increasing. The situation is exhibited in Fig. \ref{fig:RexcTh}.

\section{ Experimental aspects of inelastic nuclear excitations} 
In this section, we discuss experimental aspects of SD WIMP studies by measuring inelastic nuclear excitations. So far SD and SID WIMPs have been studied experimentally by measuring nuclear recoils of elastic scatterings. 

SD WIMPs may show fairly appreciable cross sections of inelastic spin excitations, as shown in previous sections. Experimentally, inelastic nuclear excitations provide unique opportunities
for studying SD WIMPs.  Inelastic excitations are studied by two ways,  A: singles measurement of both the nuclear recoil energy $E_R$ and the decaying $\gamma $-ray energy $E_{\gamma }$ in one detector, and B: coincidence measurement of the nuclear recoil and the $\gamma $-ray in two separate detectors. The merits of each of them are as follows.\\

{\it A: Singles measurement}
 
 The large energy signal is obtained by summing the nuclear recoil signal and the $\gamma $ ray signal. It is given as  
\begin{equation}
E(ex)=E_{\gamma} + Q(E_R(ex))E_R(ex), 
\end{equation}
where $E_R(ex)$ is the nuclear recoil energy,  $E_{\gamma }$ is the excitation
energy and $Q(E_R(ex))$ is the quenching factor for the recoil energy signal. In most scintillation and ionization detectors, the quenching factor is as small as $Q(E_R(ex)) \approx 0.1 - 0.05$. Therefore the energy deposit is mainly the excitation energy. This is much larger than just the recoil energy signal of $E(gr) =Q(E_R(gr))$, which is much quenched, depending on the detectors. 

The sharp rise of the energy spectrum at the 
energy of $E_{\gamma} + Q(E_{min}) E_{min}$, where $E_{min}$ is the minimum energy transfer to the recoil nucleus.  This makes it possible to identify the WIMP nuclear interaction.  On the other hand, the recoil energy spectrum $E_R(gr)$
is continuum like back ground at the low energy region, and thus is hard to be identified.

$E(ex)$ is well above the detector threshold $E(th)$, while
the main part of $E(gr)$ is cutoff by $E(th)$. Accordingly, the event rate $R(ex)$ is about the same order of magnitude as $R(gr)$ for SD WIMPs although the inelastic cross section is much smaller than the elastic one. \\

{\it B: coincidence measurement}\\
The nuclear recoil signal and the $\gamma $ ray signal from  adjacent two detector layers  among  a multi-layer detector array  are measured in coincidence. Here WIMP hits one layer and the $\gamma $ ray escapes from the layer and deposits the energy at the adjacent layer. The coincidence measurement reduces greatly BG counts.
The $\gamma $ ray energy signal is not quenched and is as large as 30-60 keV. It shows a sharp peak to be identified easily. 

The one layer of the detector has to be as thin as sub mm to make it possible for the $\gamma $ ray to escape from the layer. The detection efficiency is an order of 0.1 - 0.3, depending very much on the $\gamma $ ray escape probability. \\

\begin{figure}
\begin{center}
\includegraphics[width=0.85\textwidth]{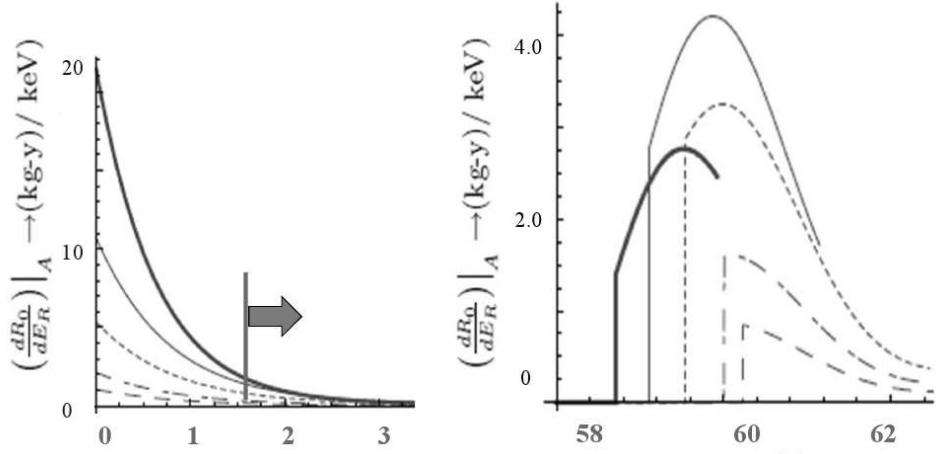}
\caption{ Energy spectrum  for WIMP $^{127}$I elastic scattering (left-hand side) and that for the 57.6 keV excited state inelastic scattering (right hand side). Quenching factor is $Q$=0.05, and the energy threshold is $E(th)$ = 1.6 keV. The x and y scales are the electron-equivalent energy and the rate per unit electron equivalent energy.}
\label{fig:I127}
\end{center}
\end{figure}

\begin{figure}
\begin{center}
\includegraphics[width=0.85\textwidth]{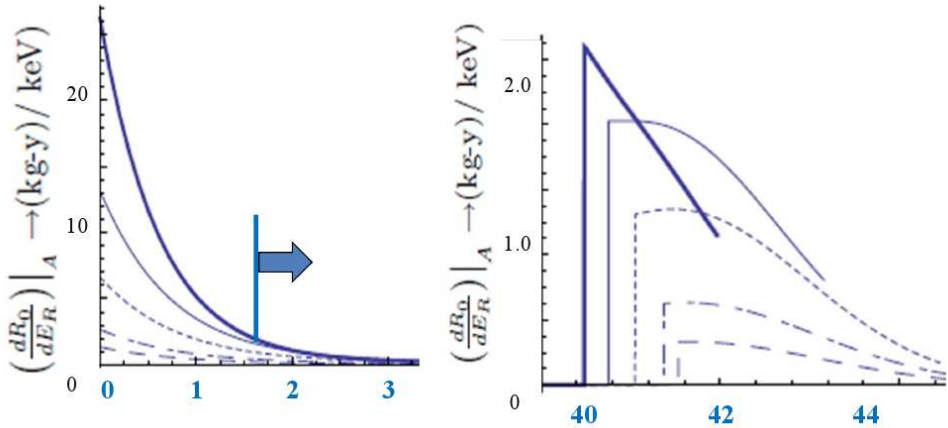}
\caption{ Energy spectrum  for WIMP $^{129}$Xe elastic scattering (left-hand side) and that for the 39.6 keV excited state inelastic scattering (right hand scale). Otherwise the notation is the same as in \ref{fig:I127}.}
\label{fig:Xe129}
\end{center}
\end{figure}

The typical energy spectra to be measured experimentally for the elastic and inelastic transitions of $^{127}$I and $^{129}$Xe are shown in Figs \ref{fig:I127} and \ref{fig:Xe129}.  Here we assumed detectors with the quenching factor of $Q$=0.05 and the energy threshold of $E(th)$ = 1.6 keV. The yield in y axis is the one per unit energy of the electron equivalent energy, i.e. $QE_R$ and the energy in x axis is the electron equivalent one. 

We note that the yield is enlarged by a factor $1/Q$=20, while the energy is shrunk by a factor $Q$=0.05. The low energy part of the elastic scatting is cut off by the thresh hold energy of 1.6 keV electron equivalent energy, i.e. 32 keV recoil energy.

\section{Concluding remarks}

 SI and SD WIMPs have extensively been studied, so far, by measuring elastic nuclear recoils.  The elastic scattering of SI WIMPs is coherent scatting, thus the cross section is enhanced by the factor $A^2$ with $A$ being the nuclear mass number.  On the other hand the elastic cross section of SD WIMPs is, is in general, smaller by 2-3 orders of magnitude than that for SI WIMPs because of lack of coherence. It may, however, compete with the coherent in models in which the spin induced nucleon cross section is much larger than the one due to a scalar interaction. We have seen that there exist viable such particle models. In such cases the inelastic WIMP-nucleus scattering  becomes important.
 
 Indeed the inelastic scattering via spin interaction provides a new opportunity for studying SD WIMPs. Experimentally, observation of both the nuclear recoils and the $\gamma $ ray following the excited state does lead to the large energy signal of the unquenched $E_{\gamma }$ and  the sharp rise of the energy spectrum at around $E_{\gamma }$.  Even though the SD inelastic cross section is smaller than the SD elastic one, the inelastic event rate is comparable with the elastic one, since the inelastic signal is well  beyond the detector threshold energy, while the elastic signal is mostly cut off by the detector threshold. 

In short, the present paper shows that the inelastic scatting opens a new powerful way to search for SD WIMPs.
  
In the present paper we discussed mainly the inelastic excitations of $^{127}$I and $^{129}$Xe by using the I and Xe detectors. Another possible isotope is the $^{73}$Ge in high energy resolution Ge detectors. We are currently evaluating the relevant nuclear matrix elements \cite{Kosmas13} (static spin and structure functions) and we will discuss this interesting case  case in a forthcoming article. 

\end{document}